\documentclass[journal=ancac3,manuscript=article,layout=traditional]{achemso}

\usepackage{graphicx}
\usepackage{amsmath}
\usepackage[version=3]{mhchem}
\usepackage[usenames]{color}
\usepackage{amssymb}

\usepackage{braket}
\usepackage{color}
\usepackage{nameref}

\usepackage{xcolor}

\usepackage[colorlinks=true, linkcolor=blue, citecolor=blue, urlcolor=blue,pdftex]{hyperref}

\setkeys{acs}{maxauthors = 20}

\usepackage{atbegshi}
\AtBeginDocument{\AtBeginShipoutNext{\AtBeginShipoutDiscard}}

\definecolor{newcolor}{rgb}{0.9,0,0.1}

\newcommand{\figref}[1]{Fig.~\ref{#1}}
\newcommand{\unit}[2]{$#1\,\text{#2}$}
\newcommand{\equnit}[3]{$#1=#2\,\text{#3}$}

\title[WS2 defects]{How Substitutional Point Defects in Two-Dimensional \ce{WS2} Induce Charge Localization, Spin-Orbit Splitting, and Strain}

\keywords{point defects, 2D materials, transition metal dichalcogenide, WS2, noncontact atomic force microscopy (nc-AFM), density functional theory (DFT), tight binding\\}

\author{Bruno Schuler}
\email{bschuler@lbl.gov}
\affiliation{Molecular Foundry, Lawrence Berkeley National Laboratory, California 94720, USA}

\author{Jun-Ho Lee}
\email{junholee@lbl.gov}
\affiliation{Molecular Foundry, Lawrence Berkeley National Laboratory, California 94720, USA}
\alsoaffiliation{Department of Physics, University of California at Berkeley, Berkeley, CA 94720, USA}

\author{Christoph Kastl}
\affiliation{Molecular Foundry, Lawrence Berkeley National Laboratory, California 94720, USA}
\alsoaffiliation{Walter-Schottky-Institut and Physik-Department, Technical University of Munich, Garching 85748, Germany}

\author{Katherine A. Cochrane}

\author{Christopher T. Chen}

\author{Sivan Refaely-Abramson}
\affiliation{Molecular Foundry, Lawrence Berkeley National Laboratory, California 94720, USA}

\author{Shengjun Yuan}
\affiliation{Key Laboratory of Artificial Micro- and Nano-structures of Ministry of Education and School of Physics and Technology, Wuhan University, Wuhan 430072, China}

\author{Edo van Veen}
\affiliation{Radboud University of Nijmegen, Institute for Molecules and Materials,
Heijendaalseweg 135, 6525 AJ Nijmegen, The Netherlands}

\author{Rafael Rold\'{a}n}
\affiliation{Instituto de Ciencia de Materiales de Madrid, ICMM-CSIC, Cantoblanco, E-28049,  Madrid, Spain}

\author{Nicholas J. Borys}
\affiliation{Department of Physics, Montana State University, Bozeman Montana 59717, USA}

\author{Roland J. Koch}
\affiliation{Advanced Light Source, Lawrence Berkeley National Laboratory, California 94720, USA}

\author{Shaul Aloni}
\author{Adam M. Schwartzberg}
\author{D. Frank Ogletree}
\affiliation{Molecular Foundry, Lawrence Berkeley National Laboratory, California 94720, USA}

\author{Jeffrey B. Neaton}
\email{jbneaton@lbl.gov}
\affiliation{Molecular Foundry, Lawrence Berkeley National Laboratory, California 94720, USA}
\alsoaffiliation{Department of Physics, University of California at Berkeley, Berkeley, CA 94720, USA}
\alsoaffiliation{Kavli Energy Nanosciences Institute at Berkeley, Berkeley, CA 94720, USA}

\author{Alexander Weber-Bargioni}
\affiliation{Molecular Foundry, Lawrence Berkeley National Laboratory, California 94720, USA}

\begin{document}
\thispagestyle{empty}


\begin{abstract}
Control of impurity concentrations in semiconducting materials is essential to device technology. Because of their intrinsic confinement, the properties of two-dimensional semiconductors such as transition metal dichalcogenides (TMDs) are more sensitive to defects than traditional bulk materials. The technological adoption of TMDs is dependent on the mitigation of deleterious defects and guided incorporation of functional foreign atoms. The first step towards impurity control is the identification of defects and assessment of their electronic properties.
Here, we present a comprehensive study of point defects in monolayer tungsten disulfide (\ce{WS2}) grown by chemical vapor deposition (CVD) using scanning tunneling microscopy/spectroscopy, CO-tip noncontact atomic force microscopy, Kelvin probe force spectroscopy, density functional theory, and tight-binding calculations. We observe four different substitutional defects: chromium (Cr$_{\text{W}}$) and molybdenum (Mo$_{\text{W}}$) at a tungsten site,  oxygen at sulfur sites in both bottom and top layers (O$_{\text{S}}$ top/bottom), as well as two negatively charged defects (CD type~I and CD type~II). Their electronic fingerprints unambiguously corroborate the defect assignment and reveal the presence or absence of in-gap defect states. 
Cr$_{\text{W}}$ forms three deep unoccupied defect states, two of which arise from spin-orbit splitting. The formation of such localized trap states for Cr$_{\text{W}}$ differs from the Mo$_{\text{W}}$ case and can be explained by their different $d$ shell energetics and local strain, which we directly measured. Utilizing a tight-binding model the electronic spectrum of the isolectronic substitutions O$_{\text{S}}$ and Cr$_{\text{W}}$ are mimicked in the limit of a zero hopping term and infinite onsite energy at a S and W site, respectively.
The abundant CDs are negatively charged, which leads to a significant band bending around the defect and a local increase of the contact potential difference. In addition, CD-rich domains larger than \unit{100}{nm} are observed causing a work function increase of \unit{1.1}{V}. While most defects are electronically isolated, we also observed hybrid states formed between Cr$_{\text{W}}$ dimers.
The important role of charge localization, spin-orbit coupling, and strain for the formation of deep defect states observed at substitutional defects in \ce{WS2} as reported here will guide future efforts of targeted defect engineering and doping of TMDs.
\end{abstract}

\maketitle
\setcounter{page}{1}

\newpage


\newpage
The introduction of impurity atoms in semiconductor crystals dramatically changes their conductivity, carrier mobility, and carrier lifetimes, among many other properties down to impurity concentrations at the parts per billion level~\cite{}. This extreme sensitivity to impurities required developing effective purification methods, and strategies to mitigate the impact of defects, for instance by passivation, and to establish precisely controlled introduction of dopants -- the foundation of modern semiconductor technology~\cite{}. 
Today, similar challenges are faced in two-dimensional (2D) materials science. Because of their intrinsic electron confinement, 2D materials are even more susceptible to structural defects~\cite{lin2016defect}.
Anticipating which defects or impurities are introduced during synthesis is often impossible given the complexity of reaction pathways and environmental variables. Moreover, predicting defect functionality, particularly for highly correlated materials, is far from trivial even with today's advanced theoretical methods and computational power. Identifying defects experimentally and probing their electronic structure is equally challenging as it requires correlating the defect's atomic and electronic structure with atomic precision and milli-electron volt energy resolution. \\

Monolayer transition metal dichalcogenides (TMDs)~\cite{novoselov20162d,manzeli20172d} are a remarkable class of 2D materials in several ways. Many are semiconducting (depending on composition and structure) with a direct band gap in the visible or near infrared range making them attractive for optoelectronic applications~\cite{mak2016photonics}. They are synthesizable by several established methods (chemical vapor deposition(CVD), metal organic chemical vapor deposition (MOCVD), atomic layer deposition (ALD), and molecular beam epitaxy (MBE))~\cite{das2015beyond}. 
They exhibit strong light-matter interactions~\cite{li2014measurement}, unique photophysics~\cite{xu2014spin,schaibley2016valleytronics}, strong spin-orbit coupling~\cite{zhu2011giant}, and prominent many-body effects due to enhanced Coulomb interactions~\cite{ugeda2014giant,yao2017optically}, exemplified by the hundredfold enhancement of exciton binding energies as compared to bulk semiconductors~\cite{Qiu2013}.\\

Tuning electronic, optical and catalytic properties of 2D materials by means of defect engineering, in particular chemical doping, is a highly anticipated technology in the field~\cite{lin2016defect}. 
In bulk semiconductors, shallow donor or acceptor states are typically used to control the carrier concentration. In 2D semiconductors, quantum confinement and screening effects generally lead to deeper defect energy levels and increasing defect ionization energies~\cite{yang2017dimensionality}. On the one hand, this limits the attainable free charge carrier concentration by chemical doping~\cite{yang2017dimensionality}. On the other hand, surface-bound deep defect levels are an ideal system for designing single-photon emitters~\cite{aharonovich2016solid,gupta2018two} and catalysts\cite{lin2016defect}.\\

Structural defects such as domain boundaries~\cite{zhou2013intrinsic,barja2016charge,wang2018atomic}, dislocations~\cite{lin2015three,wang2018atomic}, vacancies~\cite{komsa2012two,zhou2013intrinsic,klein2017robust,wang2018atomic,schuler2018large}, interstitial atoms~\cite{wang2018atomic}, antisite defects,~\cite{zhou2013intrinsic,wang2018atomic} and impurities~\cite{komsa2012two,hildebrand2014doping,gao2016transition,klein2017robust,wang2018atomic,barja2018identifying} have been previously identified in TMDs and studied theoretically~\cite{van2013grains,le2013joined,yuan2014effect,lehtinen2015atomic,komsa2015native,haldar2015systematic,li2016strong,khan2017electronic,naik2018substrate}. 
Such defects are believed to modify charge transport~\cite{qiu2013hopping}, to host defect-bound excitons,~\cite{tongay2013defects,chow2015defect,carozo2017optical,saigal2016evidence} and to act as single-photon quantum emitters~\cite{he2015single,chakraborty2015voltage,koperski2015single,srivastava2015optically,gupta2018two}. It is also suspected that point defects represent active sites in catalytic processes on the otherwise inert TMD surface~\cite{li2016all,li2016activating,hong2017atomic}.
In light of these anticipated functionalities there are growing efforts towards targeted incorporation of substitutional defects for doping and alloying of TMDs to tune their electronic, magnetic, optical, and catalytic properties.~\cite{zhang2015cvd,zhang2014two,tongay2014two,suh2018reconfiguring,wang2018atomic}\\

Establishing structure-function relationships for such defects or dopants is, however, challenging. Often macro- or mesoscopic ensemble measurements are correlated with atomic-resolution microscopy techniques to infer a certain defect functionality. The presence or prevalence of a certain defect does not necessarily indicate causality of the observed property. Instead, electronic and optical properties need to be measured locally to establish an unambiguous link to a specific structural defect. 
Moreover, two major complications impede defect identification itself by using the two prime experimental methods, aberration corrected transmission electron microscopy (AC-TEM) and scanning tunneling microscopy (STM). First, TMDs are electron beam sensitive, which can cause \textit{in-situ} generation of defects -- in particular chalcogen vacancies -- by TEM~\cite{komsa2012two,qiu2013hopping,zhou2013intrinsic,komsa2015native,hong2015exploring,wang2018atomic}. Second, identification of defect atomic structure by STM is very challenging because the STM contrast is dominated by electronic effects, which led to contradictory defect assignments by STM-only studies~\cite{liu2016point,zhang2017defect,lin2018realizing}.
Both of these complications have supported the notion that chalcogen vacancy defects are abundant in as-grown TMDs under ambient conditions, which we have recently challenged~\cite{barja2018identifying,schuler2018large}. We showed that oxygen substituting sulfur is the most abundant point defect in CVD-grown \ce{WS2}~\cite{barja2018identifying} and chalcogen vacancies, which are absent in as-grown samples, can be created by high-temperature annealing in vacuum~\cite{schuler2018large}.\\

Here, we report the direct correlation of the atomic and electronic structure of commonly observed point defects in CVD-grown monolayer \ce{WS2} samples using a combination of low-temperature scanning tunneling microscopy/spectroscopy (STM/STS), CO-tip noncontact atomic force microscopy (nc-AFM), Kelvin probe force spectroscopy (KPFS), density functional theory (DFT), and tight-binding calculations. 
Six point defects occurring in as-grown samples were found and investigated. Four of them were identified as W or S substitutions: Cr$_{\text{W}}$ (chromium substituting tungsten), Mo$_{\text{W}}$ (molybdenum substituting tungsten), and O$_{\text{S}}$ (oxygen substituting sulfur) in the top and bottom sulfur layer. In addition, we observed two types of negatively charged defects (CDs), where the impurity atom could not be unambiguously identified.\\ 

We discuss in detail the electronic defect states associated with each of these defects. Particularly the role of spin-orbit coupling, crystal-field splitting, and strain are analyzed, and the origin of the qualitatively different behavior of isoelectronic substituents are examined. One case of defect state hybridization is discussed. Furthermore, the negative charge localization at two types of defects are demonstrated, which induces significant local band bending and a dramatically higher work function of mesoscopic CD-rich domains. \\

These results advance our understanding of the role of defects in TMDs. Unambiguous chemical assignment of defects suggests formation pathways, optimized synthetic processes, and routes to controlled chemical doping. Detailed structural and electronic characterization provides insights into underlying fundamental physical principles. 
Of particular interest is the observation of spin-orbit split deep in-gap states for Cr$_\text{W}$ and the annealing-induced sulfur vacancy as well as the defect-bound charge at both CDs. The latter can be expected to form bound trions, scatter charge carriers, and act as an efficient recombination site, while the former are potential candidates for single-photon emitters. 
\\

\section*{Results and Discussion}

Monolayer \ce{WS2} was grown using a chemical vapor deposition (CVD) process on epitaxial graphene (Gr) on SiC(0001)~\cite{kobayashi2015growth,forti2017electronic,Kastl2017cvd}. The tungsten oxide (WO\textsubscript{2.9}) powder and \ce{H2S} gas precursor streams were compartmentalized to prevent sulfurization of the transition metal oxide precursor during growth~\cite{kastl2017important,Kastl2017cvd}. More details on the sample growth and preparation prior to low-temperature scanning probe inspection can be found in the Methods section and Ref.~\citenum{Kastl2017cvd}. The Gr/SiC substrate only weakly interacts with the \ce{WS2} layer preserving its interesting intrinsic band structure, such as the direct band gap in the visible range and the spin-orbit splitting at the K-point~\cite{forti2017electronic,Kastl2017cvd}. The most notable exception is a substrate-induced pinning of the \ce{WS2} Fermi level to the upper third of the band gap, near the conduction band edge~\cite{forti2017electronic,ulstrup2019nanoscale}. The reported values of the substrate-induced doping vary between different studies, most likely as a consequence of varying growth conditions, which can affect the interfacial states of the graphene/buffer layer system, for instance through intercalation.~\cite{riedl2009quasi} Also, the quasiparticle band gap becomes smaller with increasing number of graphene layers due to the increased substrate screening (see Supplementary Material)~\cite{ugeda2014giant}.\\

\figref{fig:SampleOverview}a shows a 3D representation of a STM overview image of a monolayer \ce{WS2} island on Gr/SiC. The island edges (indicated by the dashed line) are usually not clearly faceted and appear to be decorated with adsorbates but were recently also associated with air-induced oxidation of the edges~\cite{park2016scanning}. By contrast, the inside of the island is clean and a variety of point defects can be discerned. The relative frequency of different defects and the overall defect density vary considerably between different samples but both usually do not depend on the location on the sample or number of repetitions of low temperature (250$^\circ$C) vacuum annealing steps. The type of defects that are created will also depend on the synthetic conditions (such as the precursor material) and hence might vary for different growth methods. The defect density summed over all occurring defects in our samples is typically on the order of \unit{10^{12}}{cm$^{-2}$} and the defects are randomly distributed. An exception is a charged defect species, where segregated CD-rich domains were found (see Section ``\nameref{sec:CDdomain}"). 
In the two close-up images (\figref{fig:SampleOverview}b,c) five types of defects can be distinguished. Their characteristic STM topography is voltage and tip termination dependent. Here, we chose a sample bias of \unit{1.1}{V}, which is about \unit{200}{mV} above the \ce{WS2} conduction band edge where the contrast difference is most apparent. In addition, a CO-terminated tip was used for enhanced resolution~\cite{Gross2009a,Mohn2013}. The two types of charged defects can, however, not be distinguished under these scanning conditions. All discussed defects, except for the CDs, have a threefold symmetry indicative of the trigonal prismatic lattice of the host crystal. In fact, the defect's STM contrast can be used to determine the full lattice orientation, which is particularly helpful to identify 60$^\circ$ grain boundaries. 
A defect assignment by STM is, however, very challenging without \textit{a priori} knowledge of their origin, \textit{e.g.} by deliberate incorporation of a specific impurity by synthetic means. Even assigning the lattice site is not trivial given that tip- and bias-dependent contrast inversions have been observed~\cite{sanchez2010understanding,tumino2019pulsed}. These complications have led to contradictory defect assignments and have in particular supported the conception that chalcogen vacancy defects are abundant in TMDs under ambient conditions, which we could refute in two recent publications~\cite{barja2018identifying,schuler2018large}.\\

\section{\ce{WS2} Defect Assignment and Electronic Fingerprint}
\begin{figure*}[]
\includegraphics[width=\textwidth]{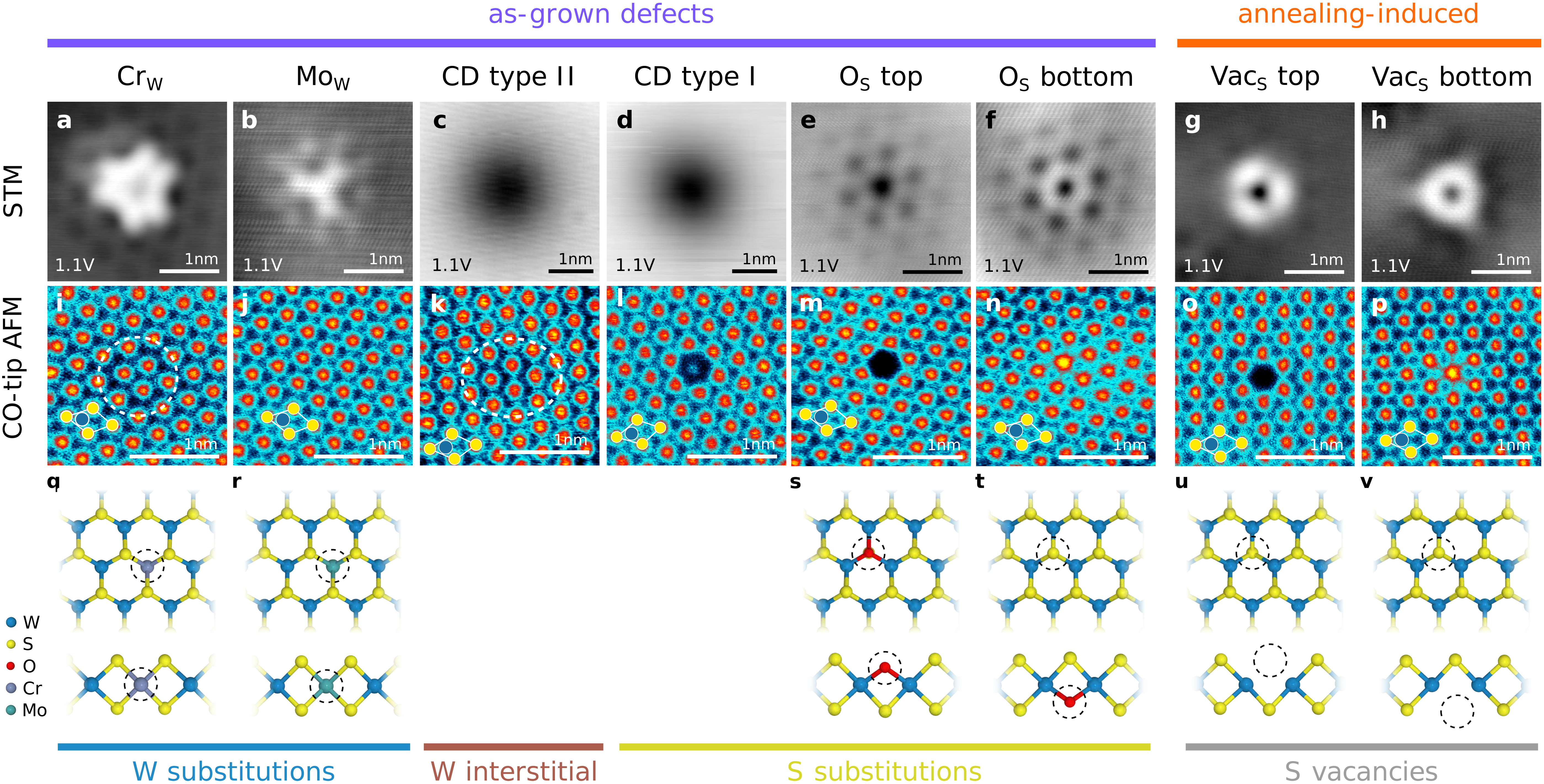}
\caption{\label{fig:AllDefects}\textbf{Point defects in \ce{WS2}.} \textbf{a-f} STM topography of the defects observed in as-grown CVD-\ce{WS2}: Cr$_\text{W}$ (a), Mo$_\text{W}$ (b), CD type~I and II (c,d) and O$_\text{S}$ top/bottom (e,f), respectively. \textbf{g,h} STM topography of annealing-induced S vacancies Vac$_\text{S}$ top/bottom. \textbf{i-p} Corresponding CO-tip AFM image of the same defects. The S and W atoms are indicated by the yellow and blue disks. The white rhombus defines the unit cell of the 1H-\ce{WS2} island. In i and k the defect location is marked by the dashed circle as a guide to the eye. \textbf{q-v} Top and side view of the DFT-calculated atomic defect structure. Defect locations are marked by a circle.
}
\end{figure*}

Here, we employ a combination of STM/STS, CO-tip nc-AFM, and DFT to provide a comprehensive characterization of all commonly observed defects in our CVD-grown monolayer \ce{WS2}: Cr$_{\text{W}}$, Mo$_{\text{W}}$, O$_{\text{S}}$ in the top and bottom sulfur sublattice, and two charged defects (CD type~I and CD type~II). In \figref{fig:AllDefects}a-f and \figref{fig:AllDefects}i-n STM topography images and the corresponding CO-tip nc-AFM images of the six \ce{WS2} defects are shown, which we find in our CVD-grown samples. A model of our defect assignment is provided in \figref{fig:AllDefects}q-t and will be justified later in this section. 
Based on the well-established contrast mechanism of CO-tip nc-AFM~\cite{Gross2009a,Moll2010,hapala2014mechanism}, the surface-layer S atoms and the lower-lying W atoms can be recognized as the circular bright and dark features in \figref{fig:AllDefects}i-p, which defines the unit cell of 1H-\ce{WS2} as indicated in the lower left corner of each panel. Assigning the W positions in the subsurface layer is important to distinguish between W substitutions and interstitial defects. 
The two rightmost columns in \figref{fig:AllDefects} depict sulfur vacancies in the top and bottom sulfur sublattice for comparison. These sulfur vacancies are \textit{not} found in as-grown samples but could be created by high-temperature annealing in vacuum~\cite{schuler2018large}.
The identification and characterization of the abundant oxygen substituent O$_{\text{S}}$ as well as the annealing-induced sulfur vacancy Vac$_\text{S}$ are discussed in detail in Ref.~\citenum{barja2018identifying} and Ref.~\citenum{schuler2018large}, respectively.
In short, the oxygen substituent at a sulfur site only marginally affects the \ce{WS2} electronic properties, in particular it does not introduce defect states in the band gap~\cite{barja2018identifying}. On the other hand, sulfur vacancies that were created by annealing in vacuum at 600$^\circ$C feature two sharp unoccupied in-gap defect states followed by vibronic satellite peaks in STS. Most importantly, the \unit{252}{meV} energy splitting between the two defect states results from a strong spin-orbit interaction~\cite{schuler2018large}. The oxygen substitution efficiently passivates the sulfur vacancy by removing its deep in-gap defect states.\\

In the following, we present the unambiguous identification of Cr$_\text{W}$ and Mo$_\text{W}$ based on their nc-AFM contrast and electronic fingerprint and provide a detailed characterization of the two charged defects, which introduce significant band bending.

\subsection{Chromium Substituting Tungsten (Cr$_\text{W}$)\label{sec:Cr_W}}
The defect assigned as Cr$_\text{W}$ (chromium substituting tungsten) (\figref{fig:AllDefects}i) is located at a W site, evident by a faint nc-AFM contrast that includes the neighboring three sulfur atoms (within encircled region). The observed contrast is in excellent agreement with nc-AFM simulations of a Cr substituent based on the probe particle model~\cite{hapala2014mechanism,hapala2014origin} (see Fig.~S6 in the Supplementary Material). In particular, the experimentally observed inward relaxation of the closest surface S atoms by 0.14 $\text{\AA}$ is quantitatively reproduced by our DFT calculations. The effect of the resulting lattice strain will be discussed in Section~``\nameref{sec:strain}". The CO-tip nc-AFM contrast is, however, not unique to Cr$_\text{W}$. A similar contrast is expected for V$_\text{W}$ (vanadium substituting tungsten), Mn$_\text{W}$ (manganese substituting tungsten), O$_\text{W}$, and S$_\text{W}$ (see Fig.~S7). However, the defect's electronic signature unambiguously identifies it as Cr$_\text{W}$ as discussed below.\\

\begin{figure}[]
\includegraphics[width=\textwidth]{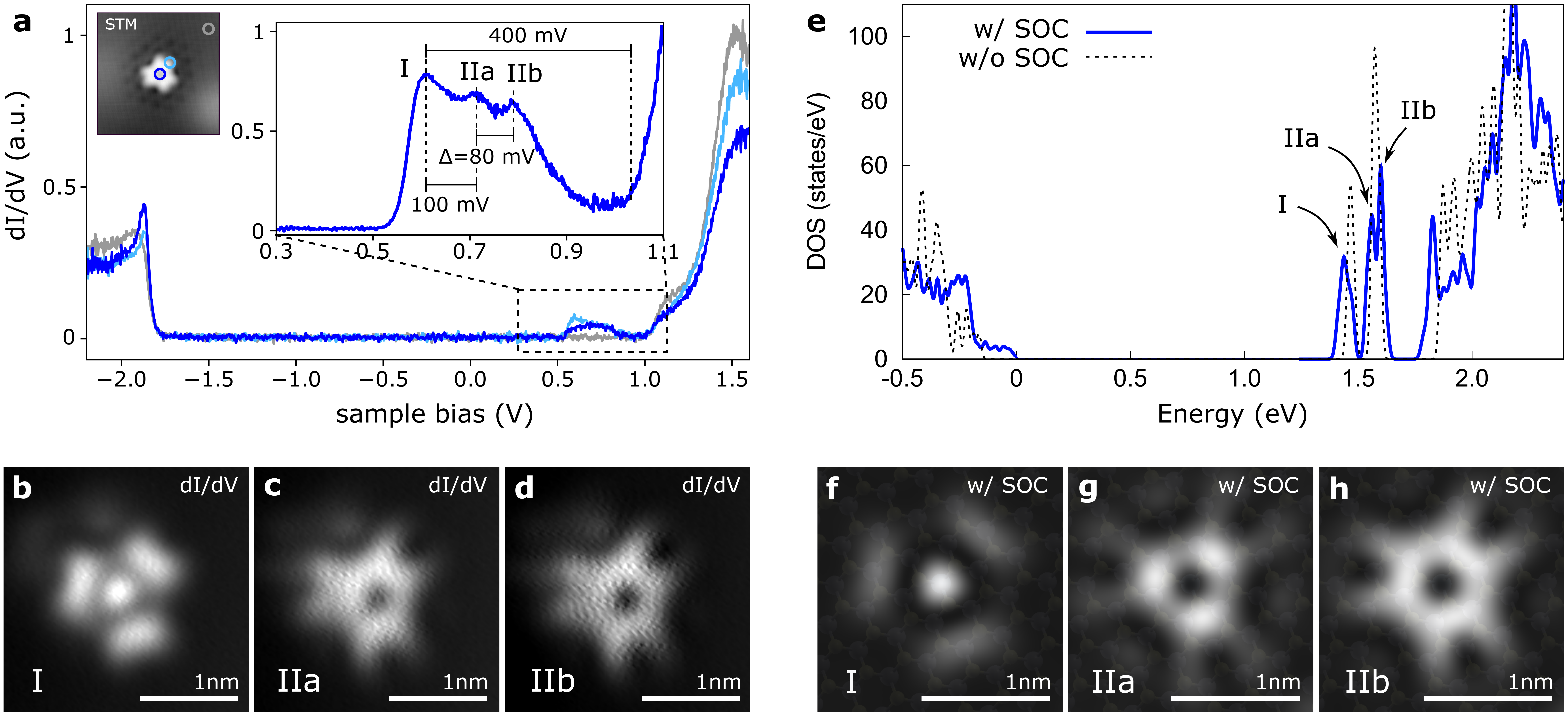}
\caption{\label{fig:Cr_W}\textbf{Defect states of Cr substituent at W site (Cr$_\text{W}$).} \textbf{a} STS measurements of a Cr$_\text{W}$ defect showing three distinct in-gap states. The circles in the inset STM image indicate the spectra positions. 
The lowest defect state I is \unit{400}{mV} below the conduction band edge and separated by \unit{100}{mV} and \unit{180}{mV} from the two consecutive defect orbitals (IIa and IIb). \textbf{b-d} Experimental dI/dV maps of the three Cr$_\text{W}$ defect states. Note that the defect orbitals IIa and IIb in c and d closely resemble each other.
\textbf{e} Calculated density of states (DOS) of \ce{WS2} containing a Cr substituent without (dashed line) and with (blue line) SOC using DFT-PBE (see Methods). The valence band maximum of the DOS with SOC is set to zero energy and the DOS without SOC is aligned such that the defect state at higher energy is located in the center of the IIa and IIb states. \textbf{f-h} Integrated local density of states for the three Cr$_\text{W}$ defect states in a $7\times7$ supercell.
}
\end{figure}

The chromium substituent can be readily observed in the STM topography as a threefold symmetric protrusion when tunneling into unoccupied states at positive sample bias (see \figref{fig:SampleOverview}b,c and \figref{fig:AllDefects}a).
In \figref{fig:Cr_W}a, a differential conductance (dI/dV) spectrum of a Cr$_\text{W}$ is shown, revealing three distinct defect resonances at \unit{400}{mV} (denoted I), \unit{300}{mV} (denoted IIa), and \unit{220}{mV} (denoted IIb) below the conduction band minimum. These resonances are indicative of localized unoccupied defect states. dI/dV images at these resonance energies displayed in \figref{fig:Cr_W}b-d map out the corresponding defect states with a spatial extent of about \unit{1}{nm}. Intriguingly, the two defect states at higher energies (denoted IIa and IIb) appear very similar, indicating a lifted orbital degeneracy.\\

The calculated DFT density of states for a Cr substituent (\figref{fig:Cr_W}e) and the shape of its associated defect orbitals (\figref{fig:Cr_W}f-h) closely resemble the experimental findings (See Methods for details of our calculations.). The close correspondence of experiment and theory corroborates our defect assignment as Cr$_\text{W}$. Other possible candidates such as V or Mn substituents can be ruled out as the former does not exhibit any in-gap state, and the defect states of the latter are distinctively different from experiment (see Fig.~S13).\\

By comparing DFT calculations with and without spin-orbit coupling (solid blue and dashed gray line in \figref{fig:Cr_W}e, respectively) we find that the $\Delta \sim$ 48 {meV} splitting between IIa and IIb is a consequence of the spin-orbit interaction. This effect will be further discussed in Section~``\nameref{sec:SOC}". 
\\

Cr substituents in CVD-grown \ce{MoS2} on \ce{SiO2} have recently also been detected by single-atom electron energy loss spectroscopy\cite{robertson2016atomic} suggesting that this substitutional transition metal atom is stable under electron-irradiation and common in TMD samples.

\subsection{Molybdenum Substituting Tungsten (Mo$_\text{W}$)\label{sec:Mo_W}}
\begin{figure*}[]
\includegraphics[width=\textwidth]{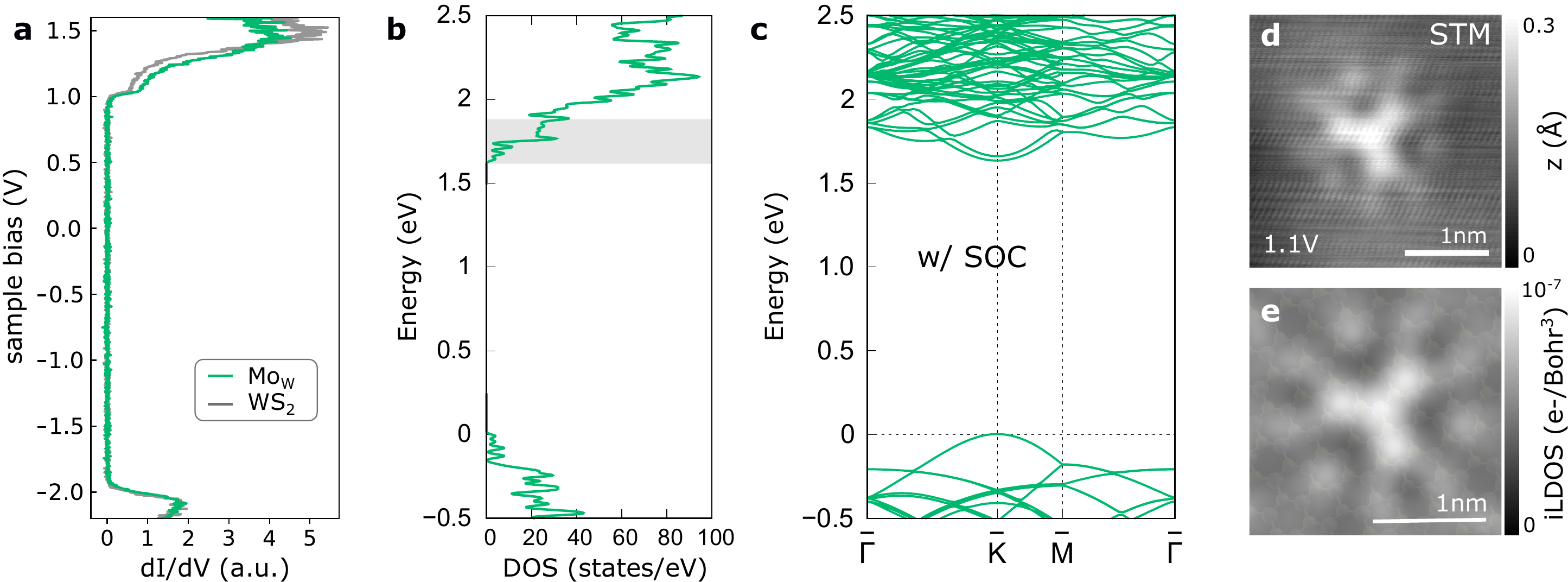}
\caption{\label{fig:Mo_W}\textbf{Electronic signature of Mo substituent at W site (Mo$_\text{W}$).} \textbf{a} dI/dV spectrum on a Mo$_\text{W}$ defect (green) and bare \ce{WS2} (gray). \textbf{b} Calculated DOS with DFT-PBE-SOC of \ce{WS2} with a Mo substituent. The valence band maximum is set to zero in energy. \textbf{c} Corresponding band structure of Mo$_\text{W}$. \textbf{d} STM topography of Mo$_\text{W}$ at \unit{1.1}{V}. \textbf{e} Integrated local density of states for an Mo$_\text{W}$ defect. The integrated energy range is marked by the gray box in b.
}
\end{figure*}

Substituting a tungsten atom by molybdenum (Mo$_\text{W}$) only minimally affects the \ce{WS2} lattice and electronic properties. In fact, Mo$_\text{W}$ does not create a detectable contrast in nc-AFM (see \figref{fig:AllDefects}j). The lattice site could only be assigned by using close-by defects as atomic markers. In this way, we could relate the center of the defect observed in STM to the W sublattice (see Fig.~S3). There is also no significant modification of the \ce{WS2} electronic spectrum evident in the dI/dV spectrum in \figref{fig:Mo_W}a.\\

The minimal effect on both the atomic and electronic structure confirms that the impurity atom is chemically similar to tungsten. Our DFT calculations indicate that indeed Mo is a likely candidate for several reasons. First, unlike the Cr$_\text{W}$ case, the host crystal lattice does not significantly distort around a Mo$_\text{W}$ impurity. The calculated in-plane sulfur-sulfur distance of the three neighboring surface S atoms only increases by \unit{3}{pm} (1\%) whereas we find a significant contraction of \unit{14}{pm} (4.5\%) for Cr (see \figref{fig:CrvsMo}). Second, the DFT band structure of \ce{WS2} with a Mo impurity, shown in \figref{fig:Mo_W}c, closely resembles the band structure of pristine monolayer \ce{WS2}. Nonetheless, slight modulations of the local density of states lead to a characteristic shape observed in STM (\figref{fig:Mo_W}d) that is nicely reproduced by the DFT integrated local density of state (ILDOS) of Mo$_\text{W}$ (\figref{fig:Mo_W}e), confirming our assignment.\\

We postulate that the source of the Cr and Mo is the \ce{WO_{2.9}} CVD precursor (99.99\%, Alfa Aesar), which contains \unit{1}{ppm} Cr and \unit{8}{ppm} Mo impurities (according to the certificate of analysis). A \unit{1}{ppm} impurity concentration translates into a doping density of \unit{2.3 \times 10^{9}}{cm$^{-2}$}. Our observed impurity density of Cr$_\text{W}$ and Mo$_\text{W}$ is on the order of \unit{10^{10}}{cm$^{-2}$}, which is plausible given that the evaporation and incorporation process of these species during CVD could be different from majority tungsten. In contrast to this dilute doping regime, other groups have deliberately synthesized ternary TMD alloys such as \ce{Mo_{x-1}W_{x}S2}~\cite{zhang2015cvd} or \ce{Mo_{x-1}W_{x}Se2}~\cite{zhang2014two,tongay2014two} to continuously tune the band gap of TMDs.\\

Consistent with our experimental results, we find that Cr and Mo impurities at a tungsten site exhibit a much lower formation energy as compared to a sulfur or interstitial site. 
For Cr$_{\text{W}}$, a formation energy of 0.31\,eV (W-rich limit) and 0.03\,eV (S-rich limit) was calculated whereas Cr$_{\text{S}}$ yields 4.72\,eV (W-rich limit) and 4.86\,eV (S-rich limit), and Cr$_{\text{i}}$ yields 6.70\,eV. These values are obtained considering \ce{WO_{2.9}} and \ce{H2S} as the source of W and S, respectively. 
In the Supplementary Material, more details on the calculation method of formation energies can be found.
For Mo$_{\text{W}}$, Mo$_{\text{S}}$ and Mo$_{\text{i}}$ we find a similar trend.
This explains why only Cr$_{\text{W}}$ and Mo$_{\text{W}}$ are thermodynamically stable and frequently observed in our samples.
We note that recent DFT calculations reveal that many transition metals at an interstitial site are energetically favorable in MoTe$_2$ where the in-plane lattice constant is $\sim$ $11\%$ larger as compared to WS$_2$~\cite{Karthikeyan2019}. In WS$_2$, the relatively small interatomic spacing renders interstitial atoms unfavorable.
We also find that O impurities at sulfur sites have a significantly lower formation energy [0.54\,eV (0.58\,eV) for W-rich (S-rich) conditions] than a sulfur vacancy [1.53\,eV (1.67\,eV) for W-rich (S-rich) conditions] in line with our observations reported in Refs.~\citenum{barja2018identifying,schuler2018large}.

\subsection{Charged Defects (CD) Type~I and Type~II\label{sec:CD}}
\begin{figure*}[]
\includegraphics[width=0.9\textwidth]{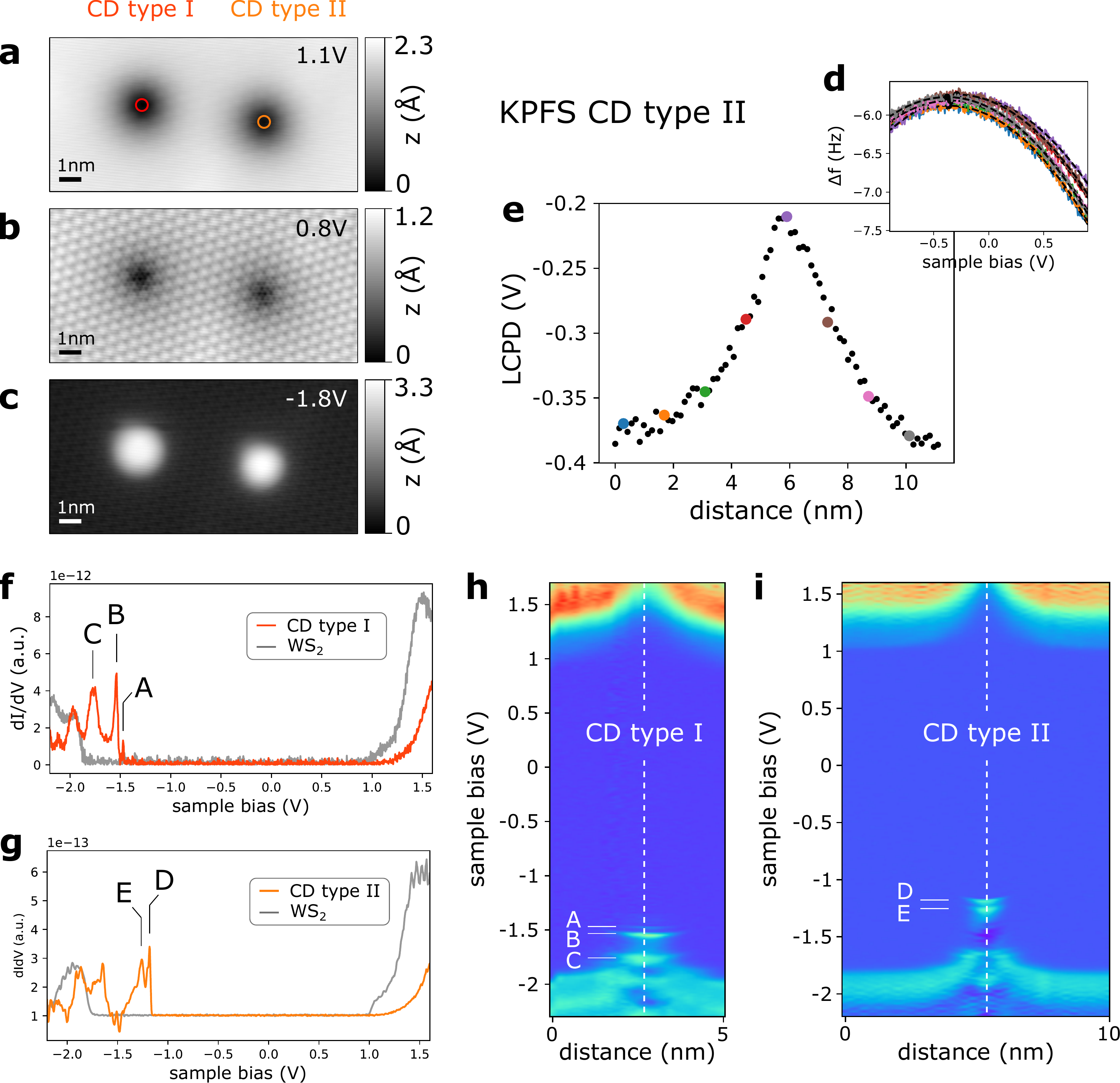}
\caption{\label{fig:CD}\textbf{Charged defect (CD) type~I and CD type~II.} 
\textbf{a-c} STM topography of the two charged defects CD type~I (left) and CD type~II (right). Both appear as a large depression at positive sample bias and a large protrusion at negative bias. \textbf{d} Kelvin probe force spectroscopy (KPFS) parabola measured across a CD type~II defect. \textbf{e} Extracted local contact potential difference (LCPD) from the vertex points in d. The colored points correspond to the subset of parabolas shown in d. The greater LCPD at the CD defect indicates negative charge. \textbf{f} dI/dV spectrum on a CD type~I (red) and on the pristine \ce{WS2} (gray). Three major resonances at negative sample bias are labelled A-C. 
\textbf{g} dI/dV spectrum on a CD type~II (orange) and on the pristine \ce{WS2} (gray). Two major resonances at negative sample bias are labelled D,E. \textbf{h,i} dI/dV spectra taken across CD type~I and CD type~II, respectively. The lateral defect positions are indicated by the vertical dashed lines. For both types of defects a clear upwards bend bending and occupied defect resonances are observed.
}
\end{figure*}

The most easily observed but most difficult to identify defect in STM is what we refer to as the charged defect (CD). In fact, we observed two types of charged defects, CD type~I and CD type~II. Both CDs appears as a large ($\sim$3\,nm) depression at positive sample bias and a protrusion at negative bias (see \figref{fig:CD}a-c). 
Similar STM contrast has been reported for synthetic \ce{MoS2}~\cite{liu2016point}, \ce{WSe2}~\cite{lin2018realizing,le2018band}, \ce{MoSe2}~\cite{le2018band} and natural bulk \ce{MoS2}(0001)~\cite{addou2015surface}. Some of these reports propose that the observed defect is negatively charged~\cite{le2018band,addou2015surface} and speculate that it might be a subsurface defect~\cite{liu2016point} or chalcogen vacancy~\cite{lin2018realizing}. This apparent ambiguity in the literature reveals again the current lack of consensus on the assignment of TMD defects.
The pronounced electronic STM contrast stems from band bending resulting from a trapped negative charge as will be discussed below. In Section~``\nameref{sec:CDdomain}", the change in work function associated with CD-rich regions will be addressed.\\

Using Kelvin probe force spectroscopy (KPFS), we could verify that both CDs are negatively charged. KPFS measures the local contact potential difference (LCPD) between tip and sample with sub-nm resolution~\cite{Mohn2012} and single electron charge sensitivity~\cite{Gross2009}. The LCPD is determined by the vertex point of $\Delta f(V)$ parabolas~\cite{Gross2009}. As seen in \figref{fig:CD}d,e the LCPD shifts about \unit{150}{mV} towards more positive bias above the CD defect, indicative of a net negative surface charge.
In STS, CD type~I (\figref{fig:CD}f,h) and CD type~II (\figref{fig:CD}g,i) exhibit a complex electronic signature with several states above the valence band maximum. Characteristic for both CDs is the significant upwards band bending of the conduction and valence band as evident from dI/dV spectra across the defects shown in \figref{fig:CD}h and \figref{fig:CD}i, respectively. This strong band bending observed within a few nanometers around the defect again suggests a negative charge trapped at the defect site. Presumably, an unoccupied acceptor state introduced by the defect gets filled from the graphene substrate, which is in tunneling contact and can exchange an integer amount of electrons. The Fermi level of graphene (\unit{0}{V} in dI/dV spectroscopy) dictates, which defect levels are filled and which are empty. In addition, the charge state of a defect will be collectively determined by
the Coulomb energy associated with the localized extra charge introduced into the system, the charge-induced lattice relaxations as well as the screening by the TMD and graphene.
\\

In \figref{fig:CD}f-i we label the major resonances of CD type~I with A-C and the major resonances of CD type~II with D and E. These electronic resonances observed at CD defects can stem from either impurity states of the substituent or 'hydrogenic' bound states in the screened Coulomb potential created by the excess negative charge~\cite{aghajanian2018tuning}. The convolution of these two effects complicates the comparison to electronic defect states obtained by DFT, which is straight-forward for neutral defects. In addition, orbital sequence reordering has been observed for charged molecular systems due to many-body interactions not captured by single-particle theories~\cite{schulz2015many,yu2017apparent}.
\\

Nevertheless we can narrow down possible CD type~I and II defect candidates. The CO-tip nc-AFM image of CD type~I shows that it is located at a sulfur site (see \figref{fig:AllDefects}l). The defect appears as a substituted atom in the top sulfur plane. Because we could remove the substituent by a voltage pulse with the tip, which left behind a sulfur top vacancy, we know that the CD type~I defect does not include a modification of the bottom sulfur layer or the tungsten layer. Unlike O$_\text{S}$ top, the substitution is clearly visible as a bright dot at the sulfur site. Two options are conceivable that would explain the observed contrast: (i) The defect could be a monoatomic substitution that is less relaxed towards the tungsten plane as compared to a oxygen substituent, but lower in vertical distance than the sulfur plane. This characteristic is predicted for C$_\text{S}$ and N$_\text{S}$. (ii) An alternative candidate is a diatomic substitution comprising a hydrogen atom, such as CH or NH. In that case, the hydrogen atom is above the sulfur plane but because of its small electron density the AFM contrast is weak. In Fig.~S8 we compare simulated CO-tip nc-AFM images for CH$_\text{S}$, NH$_\text{S}$, C$_\text{S}$, N$_\text{S}$, and O$_\text{S}$. The best match is achieved for CH$_\text{S}$, and NH$_\text{S}$.\\

In Fig.~S16 we also compare the DFT band structures of the five defect candidates mentioned above. We find that C$_\text{S}$ and N$_\text{S}$ form an unoccupied or half-occupied acceptor state, respectively. CH$_\text{S}$ features a partially-filled valence band and a resonant defect state. In all three cases, the formally unoccupied or partially occupied states are below the Fermi energy of the Gr/SiC substrate and are hence expected to be filled by the substrate. Therefore, based on the electronic spectrum, CH$_\text{S}$, C$_\text{S}$, and N$_\text{S}$ are possible candidates that generate a bound negative charge like observed at CD type~I defects. NH$_\text{S}$ and O$_\text{S}$ have fully occupied bands and are expected to be neutral.\\

Nitrogen substitutions at sulfur sites have recently been detected by electron spin resonance in synthetic bulk \ce{MoS2} samples~\cite{schoenaers2018nitrogen}. This study found that such N substituents are very stable and act as \textit{p-}type dopants, in line with our observations. Nitrogen dopants have also been deliberately introduced in \ce{MoS2} and \ce{WS2} using post-growth treatment with nitrogen plasma~\cite{azcatl2016covalent,tang2018direct}. Similarly, carbon-hydrogen (CH) units at sulfur sites could be introduced in \ce{WS2} by a methane plasma-assisted process~\cite{fu2019carbon}. \\

The identification of CD type~II is even more challenging since the CO-tip nc-AFM contrast is extremely faint as shown in \figref{fig:AllDefects}k. By adjusting the contrast and color scale, a feature can be conjectured around a W interstitial and W site, which suggests that the defect is an off-centered W interstitial. This view is supported by correlating STM and AFM images (see Fig.~S4) using an atomic O$_\text{S}$ marker similar to the procedure applied for Mo$_\text{W}$, but with the complication that the CD STM contrast is less local. Therefore, this result comes with a higher degree of uncertainty. However, we can exclude that CD type~II is a defect in the upper sulfur plane.

\section{Spin-Orbit Splitting of Cr$_\text{W}$ Defect States}\label{sec:SOC}
Spin-orbit coupling (SOC) is a defining characteristic of the TMD band structure. The combination of SOC and lack of inversion symmetry of the TMD monolayer leads to spin-momentum locking at the K point of the Brillouin zone, the key element of valleytronic concepts~\cite{mak2012control,schaibley2016valleytronics}. The effect of SOC on defect electronic states has, however, not received much attention. Recently, we measured the spin-orbit splitting between \ce{WS2} sulfur vacancy states~\cite{schuler2018large}. The exceptionally large splitting (\equnit{\Delta}{252}{mV}), in that case, is due to the W $5d$ electrons that contribute to the vacancy states.\\

\begin{figure*}[]
\includegraphics[width=0.7\textwidth]{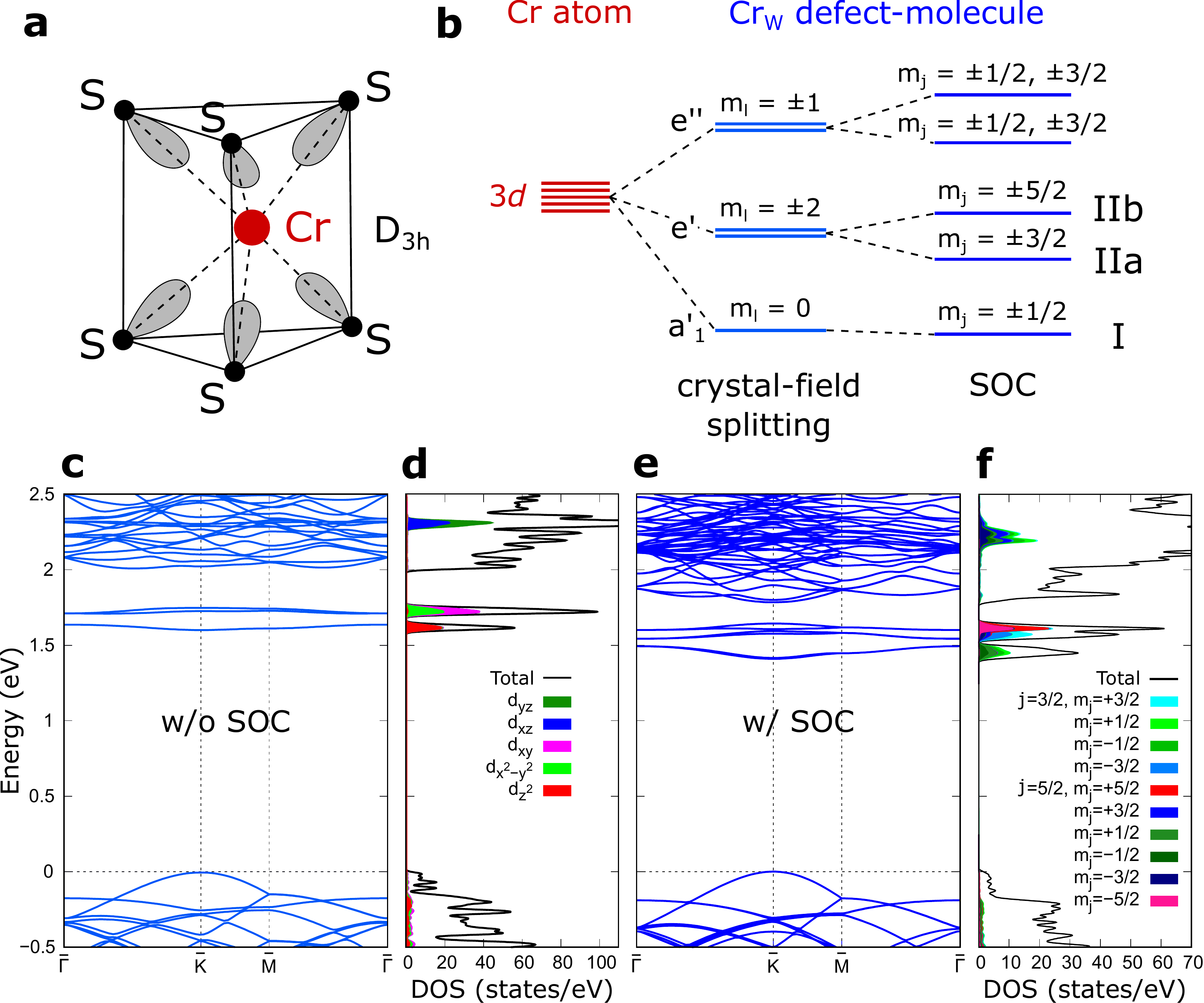}
\caption{\label{fig:SOC}\textbf{Orbital splitting of Cr$_\text{W}$ defect states.} \textbf{a} Schematic of a Cr atom in a trigonal prismatic (D$_\text{3h}$) 'ligand' field created by six neighboring S atoms. \textbf{b} Energy level diagram (not to scale) of the 'defect-molecule' model highlighting the lifting of Cr's $d$ orbital degeneracy by the crystal-field and SOC. $a_1'$, $e'$, and $e''$ denoted the irreducible representations of the D$_\text{3h}$ point symmetry group. $m_l$ and $m_j$ refer to the eigenstates of $\hat{L}_z$ and $\hat{J_z}$, respectively. \textbf{c} Band structure of \ce{WS2} with a Cr substituent without SOC. The valence band maximum is set to zero in energy. \textbf{d} Projected density of states on the Cr atomic $d$ orbitals computed with DFT-PBE. The colored area is proportional to the contribution of each $d$ orbital. \textbf{e} Band structure of \ce{WS2} with a Cr substituent with DFT-PBE-SOC. \textbf{f} Projected density of states onto eigenstates of the total angular momentum operator $\hat{J}$ labelled with the main and secondary total angular momentum quantum numbers $j$ and $m_j$, respectively. 
}
\end{figure*}

For Cr$_\text{W}$ we observe a similar effect. For Cr$_\text{W}$'s three in-gap defect states two arise from SOC, with a splitting of \equnit{\Delta}{80}{mV} (\textit{cf.} Section~``\nameref{sec:Cr_W}"). 
In a simple atomistic picture the transition metal substitution can be regarded as an atom in an effective ligand field formed by the neighboring six sulfur atoms analog to a coordination complex (see \figref{fig:SOC}a). This notion was introduced in the defect-molecule model by Coulson and Kearsley to describe color centers in diamond~\cite{coulson1957colour}. In this framework, the defect system is decoupled from the host lattice and molecular orbital theory can be applied. This approach is adequate for deep level defect states that are highly localized.\\

For the Cr $3d$ states, the initial orbital degeneracy is lifted by the crystal field splitting and spin-orbit coupling. The crystal field in the trigonal prismatic geometry splits the states into three energy levels: $d_{z^2}$, ($d_{x^2-y^2}$ / $d_{xy}$), and ($d_{yz}$ / $d_{xz}$) with increasing energy (\textit{cf.} level diagram in \figref{fig:SOC}b) and well-defined magnetic quantum numbers of $m_l = 0, \pm 2$ and $\pm 1$, respectively. These levels correspond to the irreducible representations $a_1'$, $e'$, and $e''$ of the D$_\text{3h}$ point symmetry group.
Upon including SOC the remaining degeneracy is further reduced. Because SOC is proportional to the angular momentum operator times the spin operator $\hat{L_z} \hat{S_z}$, atomic orbitals with a non-vanishing magnetic quantum number ($m_l \neq 0$) split into new eigenstates of the total angular momentum operator $\hat{J}$.\\

This qualitative picture can be quantitatively described by our fully-relativistic, noncollinear DFT calculations. The orbital character of the Cr$_\text{W}$ impurity states is obtained by projection onto atomic orbitals as shown in \figref{fig:SOC}d. As described above, we find three defect states associated with Cr $3d$ orbitals that have predominantly $d_{z^2}$, ($d_{x^2-y^2}$ / $d_{xy}$), and ($d_{yz}$ / $d_{xz}$) character, corresponding to the crystal-field splitting. Including SOC the three defect states split into five states.
Projections of the Cr$_\text{W}$ DOS with SOC onto eigenstates of the total angular momentum operator reveal that the three states in the band gap (formerly denoted I, IIa and IIb) are mainly comprised of $m_j = \pm 1/2, \pm 3/2 , \pm 5/2$ components. The two higher states in the \ce{WS2} conduction band are mixtures of  $m_j = \pm 1/2$ and $\pm 3/2$, where $j$ and $m_j$ denote the expectation value for $\hat{J}$ and $\hat{J_z}$, respectively (see \figref{fig:SOC}b,f). 
The spin-orbit splitting between IIa and IIb was calculated to be \unit{48}{meV}, which is close to experimental value of \unit{80}{meV}. \\

It should be noted that the Cr$_\text{W}$ defect states are hybrid orbitals with contributions from Cr $3d$, W $5d$, and S $3p$ states (see Fig.~S12). Hence, the intuitive atomistic view underlying the defect-molecule model is limited. However, it is justified for deep in-gap states where orbital overlap is small. The electronic character of Cr$_\text{W}$ defect states is different from that of a S vacancy that is mainly composed of W $5d$. The orbital contributions of Cr's $3d$ electrons effectively reduce the spin-orbit splitting of Cr$_\text{W}$ as compared to Vac$_\text{S}$.\\

The deep, spin-orbit split defect states observed for Cr$_\text{W}$ impurities and sulfur vacancies~\cite{schuler2018large} offer several qualities that are of potential interest for solid-state photonic spin qubits. Their discrete states are well decoupled from dispersive bulk states, they are accessible and tunable by virtue of the two-dimensionality of the host crystal, and can be coupled to electric fields \textit{via} SOC~\cite{van2018readout}. Recently, it was suggested that defect complexes of chalcogen vacancies and paramagnetic impurities such as rhenium, can act as ideal two-level systems with emission in the telecom wavelength range~\cite{gupta2018two}. Our findings for Cr$_\text{W}$ suggest that the targeted incorporation of such paramagentic impurities is feasible in a standard CVD process by controlling the impurity levels of the precursors. While chalcogen vacancies are not present in our as-grown samples~\cite{barja2018identifying}, they can be generated by thermal annealing~\cite{schuler2018large}. Alternatively, defects could be introduced site-selectively \textit{via} ion-beam irradiation~\cite{tongay2013defects,klein2019site} or atomic manipulation.

\section{Defect-Defect Interaction: Cr$_\text{W}$ Dimer}
\begin{figure*}[]
\includegraphics[width=\textwidth]{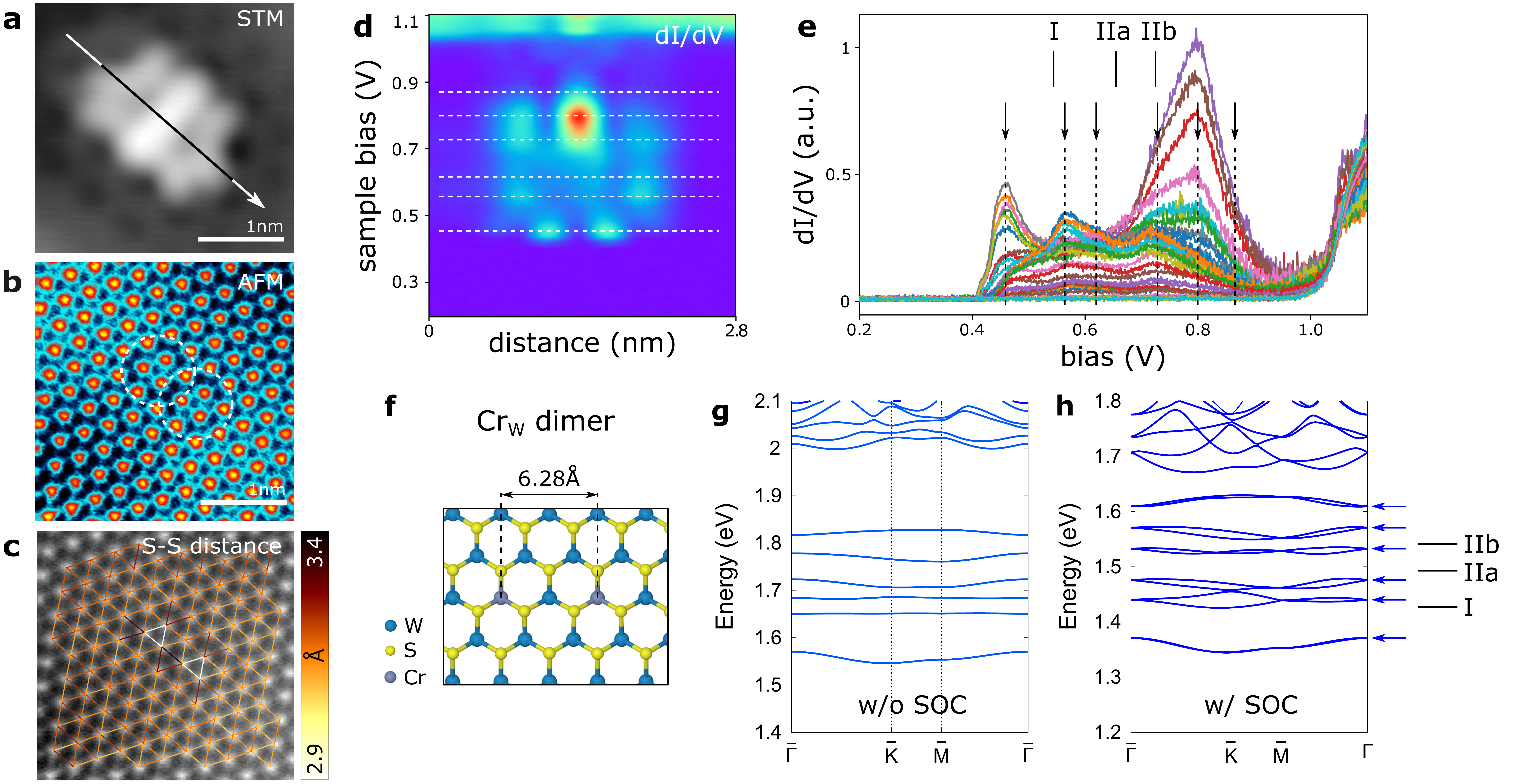}
\caption{\label{fig:DefectInteraction}\textbf{Hybrid states of a Cr$_\text{W}$ dimer.} \textbf{a} STM topography of a Cr$_\text{W}$ dimer in the third-nearest neighbor configuration. \textbf{b} CO-tip nc-AFM image of the same dimer. \textbf{c} Extracted S-S distance of the nc-AFM image shown in b revealing the local lattice strain. \textbf{d,e} dI/dV spectra recorded across the dimer (direction indicated by arrow in a). Six hybrid dimer states are marked by the dashed lines and arrows in e. The three single Cr$_\text{W}$ states (I, IIa, IIb) are indicated for comparison. \textbf{f} DFT model of the same Cr$_\text{W}$ dimer. \textbf{g,h} DFT band structure of the Cr$_\text{W}$ dimer in a 7$\times$7 supercell without (g) and with SOC (h). The six hybrid Cr$_\text{W}$ dimer states are marked with blue arrows and the three lines (I, IIa, IIb) indicate the calculated defect states of a single Cr$_\text{W}$ for comparison.
}
\end{figure*}

Defect-defect interactions become relevant as the mean defect separation becomes comparable to the dimension of the defect orbitals. In our samples we can consider most point defect as isolated. However, we find occasions where two defects are in close proximity such that defect states hybridize. In \figref{fig:DefectInteraction}, a Cr$_\text{W}$ dimer in a third-nearest neighbor configuration is shown. The dI/dV spectra across the dimer (\figref{fig:DefectInteraction}d,e) reveal a substantial hybridization between the Cr$_\text{W}$ in-gap states. 
The number of hybrid states formed is difficult to assess as the defect resonances significantly overlap. 
Six identifiable peaks are indicated by the dashed lines in \figref{fig:DefectInteraction}e. The original three Cr$_\text{W}$ states (arrows in \figref{fig:DefectInteraction}e) fall within the range of the hybrid states, consistent with the notion of bonding and anti-bonding hybrid states.\\

For the same Cr$_\text{W}$ dimer geometry we calculated the band structure with DFT in a 7$\times$7 supercell as shown in \figref{fig:DefectInteraction}g,h. For both with (\figref{fig:DefectInteraction}h) and without (\figref{fig:DefectInteraction}g) considering SOC, we find six Cr$_\text{W}$ dimer states. These non-degenerate states results from the lower crystal field symmetry as compared to an isolated Cr$_\text{W}$, and orbital hybridization between the two Cr impurities. The relative energies of these hybrid orbitals with respect to the three isolated Cr impurity states are in quantitative agreement with experiment.\\

When approaching higher Cr doping densities towards the formation of a Cr$_\text{x}$W$_\text{1-x}$S$_2$ alloy, we expect that a dispersive defect band will form in the same energy range. Defect state hybridization was only observed for two equivalent defects.
Hybrid dimers without orbital energy overlap such as Cr$_\text{W}$Mo$_\text{W}$ or Cr$_\text{W}$O$_\text{S}$ did not form mixed states.

\section{Isoelectronic Trap States Due to $d$ Shell Energetics and Strain: Cr$_\text{W}$ vs Mo$_\text{W}$}\label{sec:strain}

In Sections~``\nameref{sec:Cr_W}" and ``\nameref{sec:Mo_W}" we described the formation of deep in-gap states for chromium substituting tungsten (Cr$_\text{W}$) and their absence for molybdenum substituting tungsten (Mo$_\text{W}$). Next we discuss possible reasons for the different behavior of these isoelectronic substitutions.\\

Cr, Mo, and W are all group VI transition metals. However, unlike W, Cr and Mo are exceptions of the Aufbau principle with the following electron configurations: Cr: [Ar] 3$d^5$4$s^1$, Mo: [Kr] 4$d^5$5$s^1$, W: [Xe] 5$d^4$6$s^2$. In other aspects like their coordination chemistry Mo and W are more alike than Cr. This is reflected in the abundance of certain oxidation states. For Mo and W, the $+6$ oxidation state prevails, whereas for Cr the lower $+3$ oxidation state is most abundant. For instance, the most stable W and Mo oxides are \ce{WO3} and \ce{MO3}, whereas Cr forms \ce{Cr2O3}~\cite{Fierro2005}.\\

\begin{figure*}[]
\includegraphics[width=\textwidth]{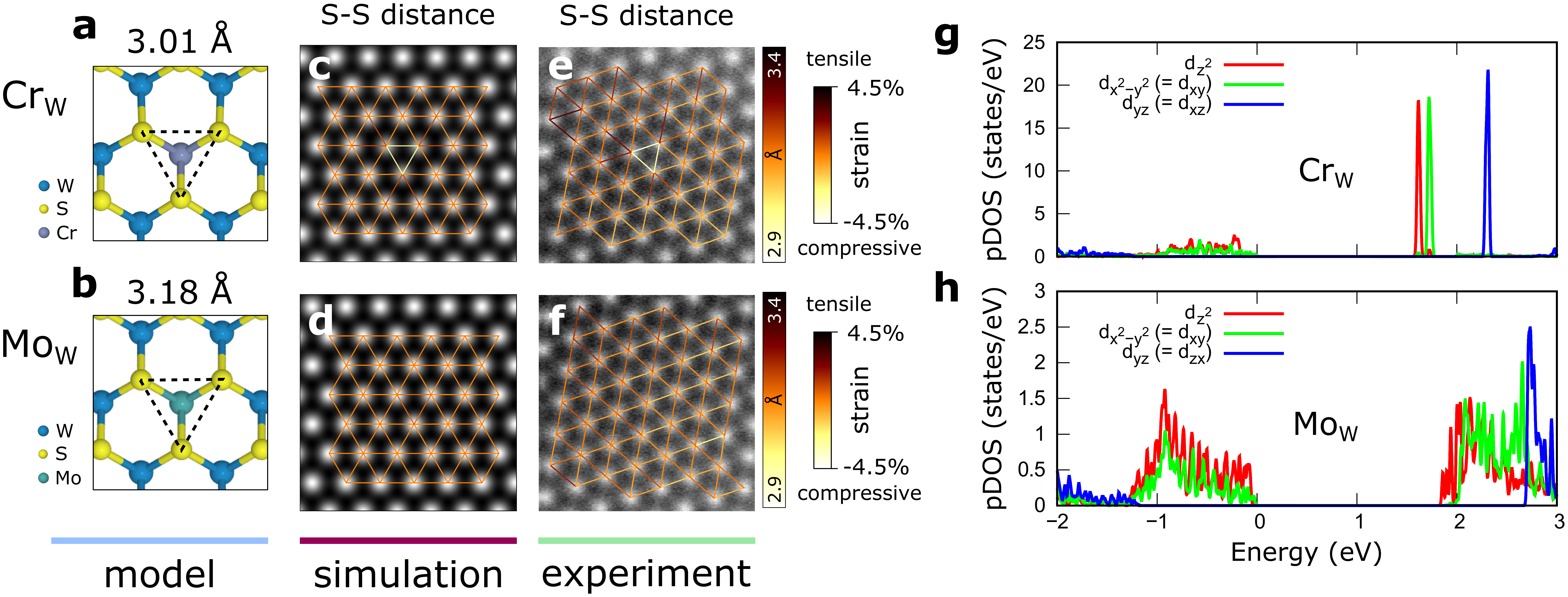}
\caption{\label{fig:CrvsMo}\textbf{Cr$_\text{W}$ vs Mo$_\text{W}$: strain and \textit{d} orbital hybridization.} \textbf{a,b} DFT-calculated relaxed defect geometry for Cr$_\text{W}$ and Mo$_\text{W}$. The dashed lines indicate the sulfur-sulfur (S-S) distance in direct proximity of the substitute. This distance is considerably smaller (\unit{3.01}{\AA}) for Cr$_\text{W}$ as compared to perfect \ce{WS2} (\unit{3.15}{\AA}), indicating compressive strain. \textbf{c,d} Simulated CO-tip nc-AFM images based on the probe particle model~\cite{hapala2014mechanism,hapala2014origin} using the defect geometries shown in a and b. \textbf{e,f} Experimental CO-tip nc-AFM image of Cr$_\text{W}$ and Mo$_\text{W}$. In c-f the atom positions have been determined using the Atomap~\cite{nord2017atomap} fitting routine revealing the local strain around Cr$_\text{W}$ and its absence around Mo$_\text{W}$ in both the simulation and experiment. The \% strain scale has been corrected for the CO tip tilting. \textbf{g,h} DFT projected DOS onto Cr and Mo $d$ orbitals showing the degree of hybridization.
}
\end{figure*}

Furthermore, the metallic radius of Cr (\unit{128}{pm}~\cite{greenwood2012chemistry}) is considerably smaller than that of both Mo (\unit{139}{pm}~\cite{greenwood2012chemistry}) and W (\unit{139}{pm}~\cite{greenwood2012chemistry}). This is reflected in the observed local strain around the Cr substituent. In \figref{fig:CrvsMo}c-f we compare the sulfur-sulfur distance in the top sulfur plane determined from the experimental nc-AFM images and nc-AFM simulations using an atom fitting routine (Atomap~\cite{nord2017atomap}). For both the experiment and simulation, we observe a reduced distance between the three surface sulfur atoms closest to the Cr substituent (white triangle in \figref{fig:CrvsMo}c,e). The apparent distance between the three sulfur atoms in the CO-tip nc-AFM image and simulation is slightly smaller than their actual geometric distance. This can be explained by the well-known CO-tip tilting~\cite{Gross2012}, which effectively enhances the relaxation effect. The simulations that are based on the probe particle model~\cite{hapala2014mechanism,hapala2014origin} account for the CO-tip tilting and yield a 6\% change in the sulfur-sulfur distance like observed experimentally. The DFT calculated defect geometry (shown in \figref{fig:CrvsMo}a) that serves as an input for the nc-AFM simulation shows a 4.5\% reduction (as mentioned in Section~``\nameref{sec:Mo_W}"). Hence, we can use the scaling factor between the DFT geometry and the simulated nc-AFM image to calibrate the effective lattice strain, indicated by the right colorbar in \figref{fig:CrvsMo}c-f. For Mo$_\text{W}$ no detectable strain was observed.\\

Lastly, the different energetics of the Cr $3d$ states and the Mo $4d$ states are discussed. 
In TMDs the states close to the valence and conduction band are made of transition metal $d$ electrons and chalcogen $p$ electrons. The degree of hybridization and the energetic location of the hybrid states depend on the level alignment between these $d$ and $p$ electrons. In \figref{fig:CrvsMo}g,h we plot the projected density of states of \ce{WS2} with a W substitution onto the substituent's $d$ orbitals. We find that for Mo$_\text{W}$ the valence band and conduction band edge are comprised of Mo's $d_{z^2}$, $d_{x^2-y^2}$, and $d_{xy}$ states, as in the non-defective case. For Cr$_\text{W}$ in contrast, $d_{z^2}$, $d_{x^2-y^2}$, and $d_{xy}$ contribute to a lesser extent to the valence band but completely unmix in the unoccupied density of states, creating the Cr$_\text{W}$ in-gap defect states, as discussed in detail in Section~``\nameref{sec:SOC}".\\

The qualitatively different behavior of the isoelectronic substitutes Cr and Mo in \ce{WS2}, in particular the isoelectronic trap states introduced by Cr$_\text{W}$, can therefore be rationalized by two main effects: Cr's smaller atomic radius that leads to local strain and the different energetics of $3d$ vs $4d$ orbitals. These differences are similarly reflected in the different coordination chemistry of group VI transition metals.\\

\section{Tight-Binding Model for Localizing Isoelectronic Substitutes}
\begin{figure*}[]
\begin{center}
\includegraphics[width=0.7\textwidth]{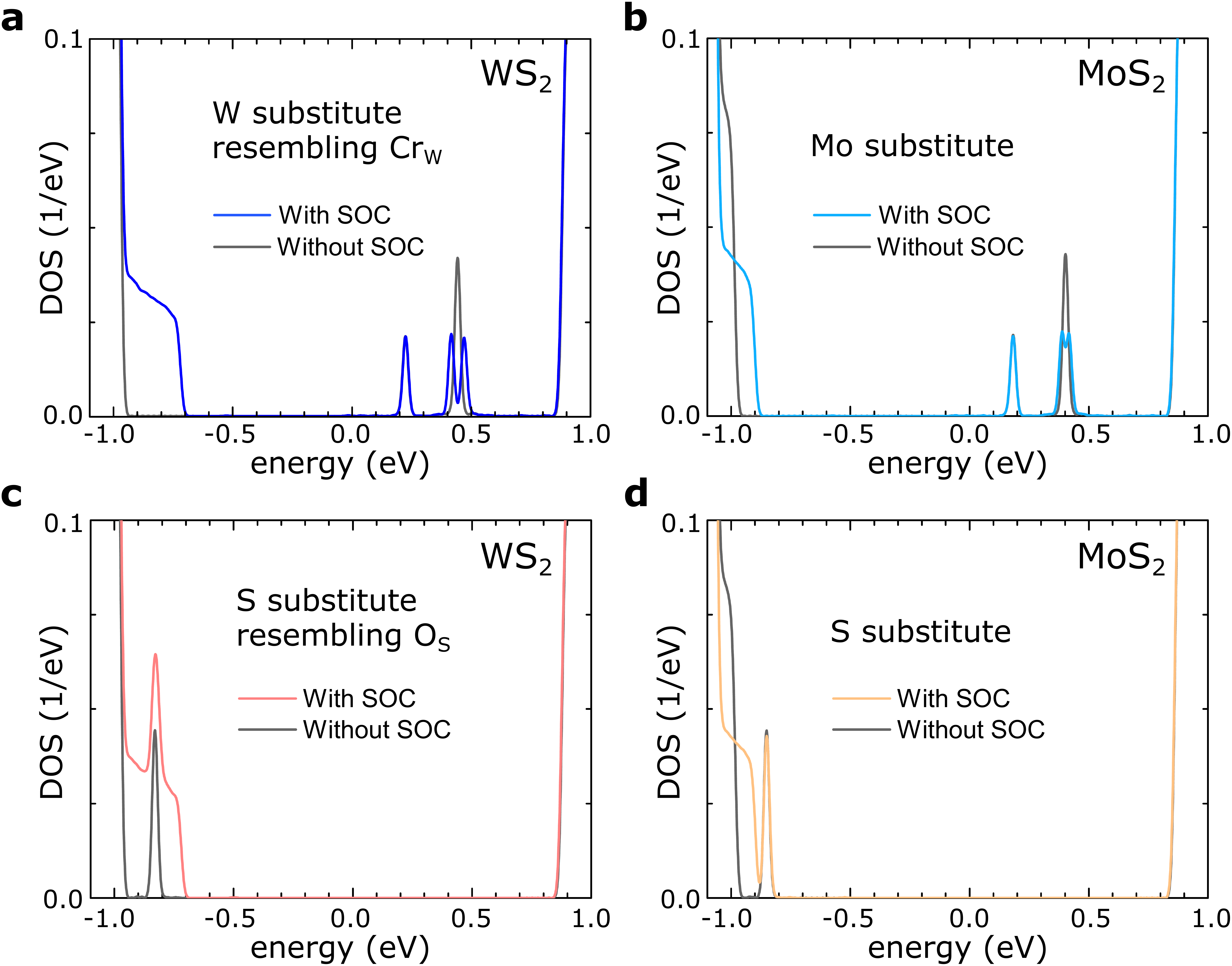}
\end{center}
\caption{\label{fig:TightBinding}\textbf{Tight-binding model for localizing transition metal and chalcogen substitutions.} \textbf{a,b} Calculated DOS for a localizing W substitution and S substitution in WS$_2$ with and without SOC. \textbf{e,f} Calculated DOS for a localizing Mo substitution and S substitution in MoS$_2$ with and without SOC. }
\end{figure*}

Above we have shown that DFT calculations can accurately describe structural and electronic properties of point defects in monolayer \ce{WS2}. We now compare our DFT results to tight-binding (TB) models that have been used as an alternative approach to describe TMD electronic properties. As we discussed earlier, the $d_{xz}$ and $d_{yz}$ orbitals are pushed far away from the band gap region due to crystal-field splitting in the D$_\text{3h}$ point group symmetry\cite{li2016strong}. Thus, the basic electronic characteristics of monolayer TMD such as the direct band gap at K point and the spin-valley coupling can be obtained by considering a minimal basis consisting of the three transition metal $d$ orbitals ($d_{xy}$, $d_{x^2-y^2}$, and $d_{z^2}$) that have even parity with respect to the $\sigma_h$ symmetry plane of the crystal~\cite{xiao2012coupled}. For point defects, which in general reduce the symmetry of the crystal structure, induce local atomic distortions, and create a different chemical environment, the basis set has to be extended to include also $d_{xz}$ and $d_{yz}$ orbitals in order to predict the correct energy spectrum. For instance, the electronic structure of monolayer TMDs with transition metal or chalcogen vacancies has been calculated in the framework of a 6-band TB model, which differs from that obtained by a 11-band TB model~\cite{yuan2014effect,khan2017electronic}. The latter can describe the electronic structure of actual vacancies and gives a good agreement with our DFT calculations for both $\textit{actual}$ W and S vacancies (see Fig.~S14 and S15). Intriguingly, the electronic structure of W and S vacancies in the framework of the $\textit{6-band}$ TB model resembles DFT calculations for Cr$_\text{W}$ and O$_\text{S}$ very well (see \figref{fig:TightBinding}), meaning that the vacancies in the 6-band TB model effectively represent the electronic spectrum of isoelectronic substitutions as opposed to actual vacancy defects. The vacancies in the 6-band TB model can be regarded as the limiting case of a zero hopping integral and infinite onsite energy at the defect site. This suggests that Cr and O substituents strongly localize electrons in the vicinity of their atomic site.

This may come from the fact that covalently bonded isoelectronic substitutes restore the crystal structure to a certain extent, such that the Hamiltonian can be approximately decoupled into even and odd parts, just like for pristine TMDs. Hence, a smaller basis set of just even orbitals may suffice.
Depending on the type of substitute, the degree of hybridization with its neighboring atoms can vary substantially. Mo$_\text{W}$ represents a case of strong hybridization (supporting extended Bloch states), which essentially restores the \ce{WS2} band structure (\textit{cf.} Section~``\nameref{sec:Mo_W}"). 
Cr$_\text{W}$ and O$_\text{S}$ represent cases of strong electron localization, simulated by describing them as 'vacancies' in the reduced 6-band TB model. \\

\section{Work Function Change at CD-Rich Domains}\label{sec:CDdomain}

\begin{figure*}[]
\includegraphics[width=0.7\textwidth]{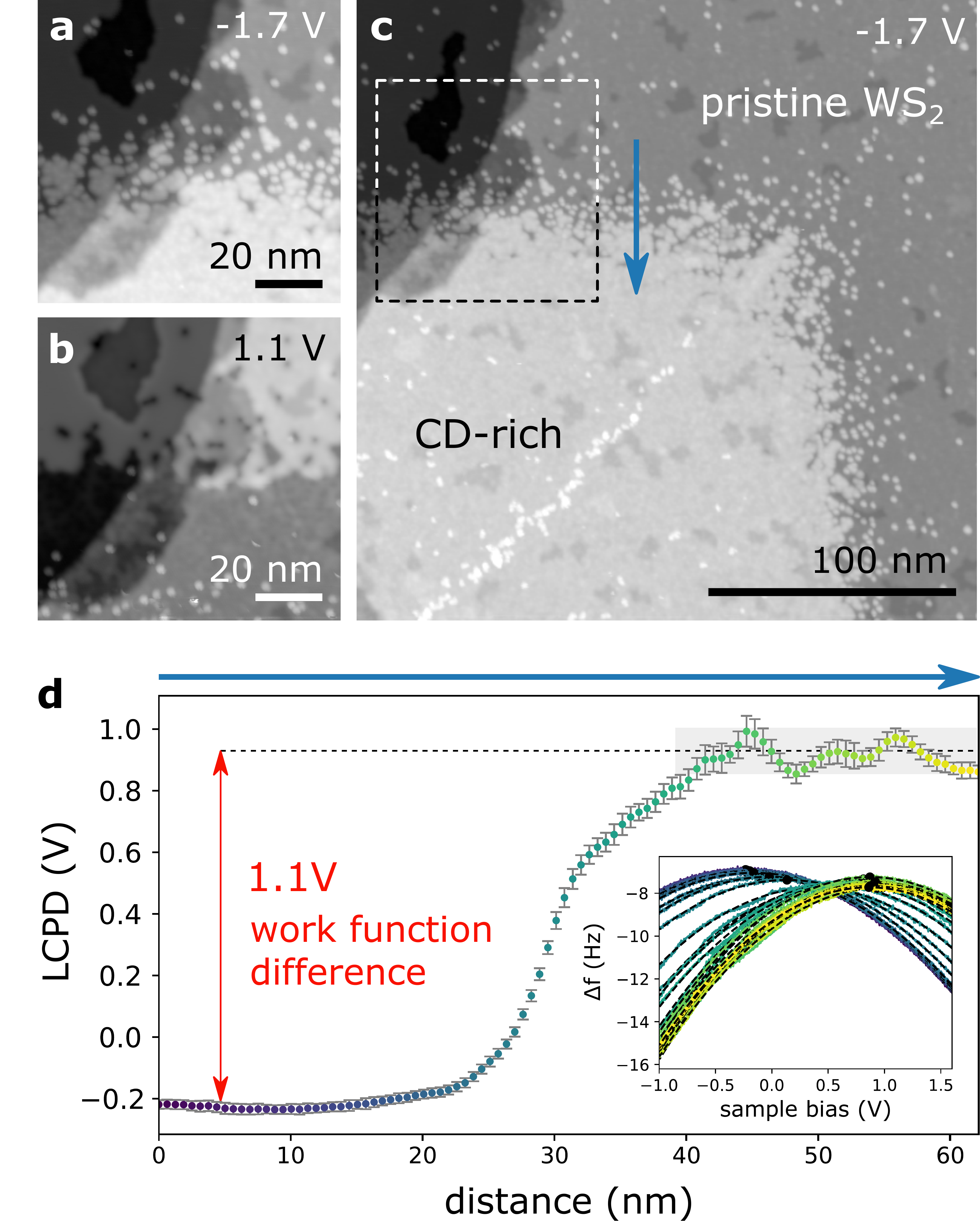}
\caption{\label{fig:charge}\textbf{Large work function shift in CD-rich domain.} \textbf{a,b} STM topography at \unit{-1.7}{V} and \unit{1.1}{V}, respectively, of a domain boundary separating pristine \ce{WS2} from a highly defective region with many CD defects (bottom part). The bias dependent contrast inversion is also observed for a single CD. \textbf{c} Larger STM overview image of the same \ce{WS2} island. The dashed box indicates the scan area displayed in a and b. \textbf{d} Kelvin probe force spectroscopy (KPFS) across the CD-rich domain boundary indicated by the blue arrow in c. The shift in the local contact potential difference (LCPD) reveals the large work function difference of \unit{1.1}{V} between the pristine \ce{WS2} and the CD-rich area. The inset shows a subset of KPFM parabolas from which the LCPD was extracted.
}
\end{figure*}

While most of the charged defects (CDs) are usually randomly distributed (like all other defects), in some samples we observe CD-rich domains with a very high CD density. In \figref{fig:charge}a-c, a CD segregated domain is shown. Such domains are at least a few \unit{100}{nm} in size. The transition from a pristine \ce{WS2} area with a normal defect density to the highly defective CD-rich area occurs over about \unit{20}{nm}. \figref{fig:charge}a,b display such a transition area. Just as for a single CD we find a bias-dependent contrast inversion originating from the upwards band bending (\textit{cf.} Section~``\nameref{sec:CD}"). \\

Using KPFS we study the work function difference between the pristine and highly defective regions. KPFS measurements across a domain boundary (blue arrow in \figref{fig:charge}c) are shown in \figref{fig:charge}d. While on the pristine \ce{WS2} an LCPD of about \unit{-0.2}{V} is measured, the LCPD is significantly increased to about \unit{0.9}{V} in the CD-rich region. This striking increase of the work function by \unit{1.1}{V} at the CD-rich domain, is a consequence of the bound negative charge at CD defects.\\

The reason for the formation of CD-rich domains in certain samples is not known. The segregation of CD defects might be thermodynamically favorable at higher CD densities, or extended domains only form in a certain growth temperature regime. So far unexplored is also whether these domains are located in a specific location of \ce{WS2} islands, such as island edges or centers~\cite{bao2015visualizing}.\\

\section*{Conclusions}
In summary, we identified and characterized the observed point defects in CVD-grown monolayer \ce{WS2} using a combination of low-temperature STM/STS, CO-tip nc-AFM, Kelvin probe force spectroscopy, and DFT calculations. 
The defects' characteristic electronic fingerprint enables an unambiguous defect assignment and reveals their optoelectronic functionality.
We found two tungsten substitutions: Cr$_\text{W}$, Mo$_\text{W}$, and three sulfur substitutions: O$_\text{S}$ in the top and bottom sulfur layer and one type of charged defect (CD). Sulfur vacancies were not observed in as-grown samples but could be generated by high-temperature annealing in vacuum.~\cite{schuler2018large}\\

Cr$_\text{W}$ forms three deep, unoccupied defect states. Two of these states arise from spin-orbit coupling as observed by STM orbital imaging and corroborated by DFT calculations. The formation of isoelectronic trap states of Cr$_\text{W}$ is in contrast to Mo$_\text{W}$, which does not form localized in-gap states. This strikingly different behavior results from the different energetics of Cr's $3d$ vs Mo's $4d$ electrons as well as lattice strain. The local strain field was directly mapped by CO-tip nc-AFM images.
In addition, the defect electronic structure of both Cr$_\text{W}$ and O$_\text{S}$ can be explained by a stronger binding of electrons at these isoelectronic substituents as described by a simple tight binding model. While most other defects are randomly distributed and can be considered isolated, we find some cases of interacting defects that form hybrid defect orbitals as exemplified by a Cr$_\text{W}$ dimer.\\

At two types of CDs, we observe a strong upwards bend bending and increase of the local contact potential difference indicative of the localization of negative charge. Some samples were found to feature CD-rich domains of a few hundred nanometers in size that lead to a significant work function increase.  
\\

The deep in-gap defect states of Cr$_\text{W}$ and the negative charge localization at CDs are highly relevant for the transport and optical properties of \ce{WS2}. Both will act as effective radiative recombination centers and might host defect-bound excitons. The negatively charged CDs are also expected to strongly scatter charge carriers and potentially trap excitons to form localized trions. The defect properties reported here will guide future efforts of targeted defect engineering and doping of TMDs.

\section*{Methods}
{\small
\subsection*{\ce{WS2}/MLG/SiC Growth Procedure}
\ce{WS2} few-layer islands were grown by a modified chemical vapor deposition process~\cite{Kastl2017cvd} on graphitized (6H)-SiC substrates~\cite{emtsev2009towards}. We used WO\textsubscript{2.9} powder (99.99\%, Alfa Aesar) and \ce{H2S} gas as the metal and chalcogen precursors, respectively. The samples were not deliberately doped. The growth temperature was 900\,$^\circ$C and the growth time was \unit{1}{h}. Further details can be found in the Supplementary Material.

\subsection*{Scanning Probe Measurements}
\paragraph{Setup.}
The experiments were performed using a CreaTec low-temperature ($T \approx 6\,\text{K}$), ultra-high vacuum ($p\approx 10^{-10}\,\text{mbar}$) combined STM and AFM. The sensor was based on a qPlus~\cite{Giessibl1999} quartz-crystal cantilever design operated in the frequency-modulation mode~\cite{Albrecht1991} (resonance frequency $f_{0}\approx 30\,\text{kHz}$, spring constant $k\approx 1800\,\text{N/m}$, quality factor $Q\approx 30,000$, and oscillation amplitude $A\approx 1\,\text{\AA}$). The voltage was applied to the sample. STM images were taken in constant-current mode at \equnit{I}{10-100}{pA} and \equnit{V}{1.1}{V} if not stated otherwise. STS spectra are performed in constant-height mode with a lock-in amplifier with a frequency of \unit{670}{Hz} and \unit{5}{mV} amplitude. The STS spectra have been characterized on Au(111) to ensure a flat tip density of states.
AFM measurements were acquired in constant-height mode at \equnit{V}{0}{V}. \\

\paragraph{Sample and tip preparation.}
The CVD grown \ce{WS2}/Gr/SiC was annealed \textit{in vacuo} at about \unit{250}{$^\circ$C} for \unit{30}{min}. We used a focused ion beam cut PtIr tips or chemically etched W tips that were sharpened by repeated indentations into a Au substrate. A CO tip was created by picking-up a single CO molecule from the Au(111) surface~\cite{Gross2009a,Mohn2013}. Both the Au(111) and the \ce{WS2}/Gr/SiC were mounted on the same sample holder.

\subsection*{Density-Functional Theory (DFT) Calculations}
We performed first-principles DFT calculations using \textsc{quantum-espresso}~\cite{Giannozzi2009,Giannozzi2017}. We used the PBE generalized gradient approximation~\cite{Perdew1996} for the exchange-correlation functional with scalar and fully relativistic optimized norm-conserving Vanderbilt (ONCV) pseudopotentials from PseudoDojo library~\cite{Hamann2013}.  We used the experimental lattice parameter of 3.15\,$\text{\AA}$~\cite{schutte1987crystal} and used a $5\times5$ or $7\times7$ \ce{WS2} supercell with $\sim$ 15\,$\text{\AA}$ vacuum region. We applied plane wave cutoff energy of 100 Ry and $4\times4\times1$ (3$\times$3) $k$-point sampling for 5$\times$5 (7$\times$7) supercell. We simulated the orbital character by means of integrated local density of states (ILDOS). More details on the effect of DFT functional and lattice constant can be found in the Supplementary Material.

\subsection*{Tight-Binding Model}
The tight-binding calculations are implemented following Ref.~\citenum{yuan2014effect}. We consider single layers of WS$_2$ containing $2400\times 2400$ atoms, with periodic boundary conditions. The electronic band structure is considered from a tight-binding model that contains six-bands: three W 5$d$ orbitals ($d_{xy}$, $d_{x^2-y^2}$, and $d_{z^2}$) and three S 3$p$ orbitals ($p_x$, $p_y$, and $p_z$)~\cite{Cappelluti_PRB_2013}. Intra-atomic spin-orbit coupling is included in the model.
The density of states (DOS) and quasi-eigenstates are obtained numerically from the Tight-Binding Propagation Method (TBPM)~\cite{Yuan_2010}.
In the simulations, localizing substitutions are mimicked by randomly removing sulfur or tungsten atoms without lattice relaxation. More details on the tight-binding model can be found in the Supplementary Material. \\

}

\paragraph{Acknowledgments}
We would like to thank Thomas Seyller for help preparing the graphene on SiC substrates and Prokop Hapala for support with setting up the AFM simulations. 
This work was performed at the Molecular Foundry supported by the Office of Science, Office of Basic Energy Sciences, of the U.S. Department of Energy under Contract No. DE-AC02-05CH11231. B.S. appreciates support from the Swiss National Science Foundation under project number P2SKP2\_171770. 
J.-H.L. and J.B.N. were supported by the Theory FWP, which is funded by the Department of Energy, Office of Science, Basic Energy Sciences, Materials Sciences and Engineering Division, under Contract No. DE-AC02-05CH11231. C. K. gratefully acknowledges support by the Bavaria California Technology Center (BaCaTeC) and the International Graduate School of Science and Engineering (IGSSE) through project ``CommOnChi''. A.W.-B. was supported by the U.S. Department of Energy Early Career Award.
S. Y. acknowledge the financial support by the National Key R\&D Program of China (Grant No. 2018FYA0305800) and computational resource provided by the Supercomputing Center of Wuhan University.

\paragraph{Supporting Information Available:}
CVD growth details of \ce{WS2} on Gr/SiC, lattice site identification for Mo$_\text{W}$, and CD type II, CO-tip nc-AFM simulations of \ce{WS2} defects, DFT calculations of other substitutional defects considered, details on the tight-binding calculations. This material is available free of charge \textit{via} the Internet at http://pubs.acs.org.

\bibliography{WS2DefectsOverview}

\clearpage
\section*{Figures}

\begin{figure*}[]
\includegraphics[width=0.6\textwidth]{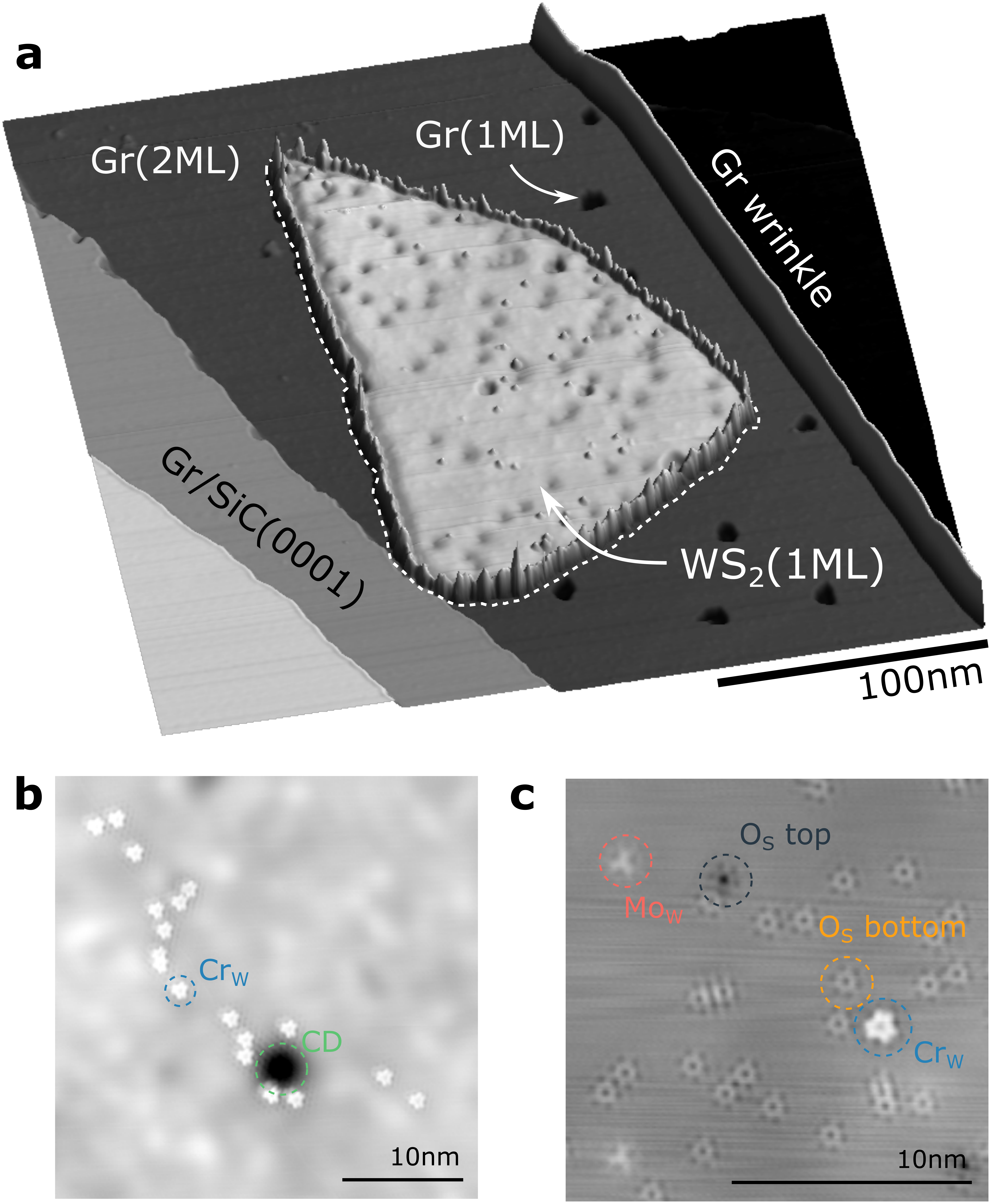}
\caption{\label{fig:SampleOverview}
\textbf{Monolayer \ce{WS2}/Gr/SiC sample.} \textbf{a} Pseudo 3D STM topography of a \ce{WS2} island on Gr/SiC. \textbf{b,c} STM images (\equnit{I}{20}{pA},\equnit{V}{1.1}{V}) of single defects on \ce{WS2}. One of each defect type is labelled. }
\end{figure*}

\clearpage

\end{document}